\pdfoutput=1

\documentclass[11pt,twoside,a4paper,cmspaper,final,collab]{cms-tdr}
\def\svnVersion{327233}\def\cmsCernNo{2013-037}\def\cmsCernDate{\today}\def\cmsMessage{Published in Physical Review Letters as \href{http://dx.doi.org/10.1103/PhysRevLett.116.052002}{\doi{10.1103/PhysRevLett.116.052002.}}}

\begin{document}\cmsNoteHeader{TOP-15-003}

\hyphenation{had-ron-i-za-tion}
\hyphenation{cal-or-i-me-ter}
\hyphenation{de-vices}

\RCS$Revision: 326442 $
\RCS$HeadURL: svn+ssh://svn.cern.ch/reps/tdr2/papers/TOP-15-003/trunk/TOP-15-003.tex $
\RCS$Id: TOP-15-003.tex 326442 2016-02-18 11:26:35Z grohsjea $

\ifthenelse{\boolean{cms@external}}{\providecommand{\cmsLeft}{top\xspace}}{\providecommand{\cmsLeft}{left\xspace}}
\ifthenelse{\boolean{cms@external}}{\providecommand{\cmsRight}{bottom\xspace}}{\providecommand{\cmsRight}{right\xspace}}
\ifthenelse{\boolean{cms@external}}
{\providecommand{\suppMaterial}{the supplemental material
  [URL will be inserted by publisher]}}
{\providecommand{\suppMaterial}{Appendix~\ref{app:supp_material}}}

\providecommand{\eepm}{\ensuremath{\Pep\Pem}\xspace}
\providecommand{\mmpm}{\ensuremath{\Pgmp \Pgmm}\xspace}
\providecommand{\empm}{\ensuremath{\Pe^\pm \Pgm^\mp}\xspace}
\providecommand{\VV}{\ensuremath{\cmsSymbolFace{VV}}\xspace}

\providecommand{\amcatnlo}{\textsc{mg5\_amc@nlo}\xspace}
\providecommand{\madspin}{\textsc{madspin}\xspace}
\newcommand{\xsec}{\ensuremath{\sigma_{\ttbar} = 746 \pm 58\stat \pm 53\syst\pm 36\lum\unit{pb}}\xspace}
\newcommand{\relerr}{12}
\newcommand{\usedLumi}{43\pbinv}

\cmsNoteHeader{TOP-15-003}
\title{Measurement of the top quark pair production cross section in proton-proton collisions at \texorpdfstring{$\sqrt{s}=13$\TeV}{sqrt(s)=13 TeV}}

\date{\today}

\abstract{
The top quark pair production cross section is measured for the first
time in proton-proton collisions at $\sqrt{s} = 13\TeV$ by the
CMS experiment at the CERN LHC, using data corresponding to an
integrated luminosity of 43\pbinv. The measurement is
performed by analyzing events with at least one electron and one muon
of opposite charge, and at least two jets. The  measured cross section
is $746 \pm 58\stat \pm 53\syst\pm 36\lum\unit{pb}$, in agreement with the expectation from the
standard model.
}

\hypersetup{%
pdfauthor={CMS Collaboration},%
pdftitle={Measurement of the top quark pair production cross section in proton-proton collisions at sqrt(s)=13 TeV}%
pdfsubject={CMS},%
pdfkeywords={CMS, physics, top physics}}

\maketitle

The measurement of \ttbar production at a center-of-mass energy not
previously accessed has great discovery potential for physics beyond
the standard model (SM), because new phenomena can significantly
enhance the \ttbar cross section. The increased energy also allows
for a test of the production mechanism, dominated at the CERN LHC by
gluon-gluon fusion, and of the validity of the theory of quantum
chromodynamics (QCD). Furthermore, top quark production is an
important source of background in many searches for physics beyond the
SM, and its accurate evaluation is important. Previously, large
samples of top quark events were collected in proton-proton collisions
at the LHC at $\sqrt{s}=7$ and 8\TeV and used to study \ttbar
production in different final states by the
ATLAS~\cite{ATLAStt1,ATLAStt2,ATLAStt3,ATLAStt4,ATLAStt5,ATLAStt6,ATLAStt8,ATLAStt9,ATLAStt10,ATLAStt11,ATLAStt12}
and
CMS~\cite{CMStt1,CMStt2,CMStt3,CMStt4,CMStt5,CMStt6,CMStt7,CMStt8,CMStt9}
collaborations.

This letter presents the first measurement of the \ttbar production
cross section $\sigma_{\ttbar}$ at $\sqrt{s}=13\TeV$, utilizing data
corresponding to an integrated luminosity of \usedLumi recorded by
the CMS experiment. In the SM, top quarks are produced predominantly
in \ttbar pairs via the strong interaction, and each top quark decays
almost exclusively to a \PW\ boson and a \cPqb~quark. For this study,
we select events that contain at least one electron and one muon of
opposite charge, and at least two jets.

The central feature of the CMS detector~\cite{Chatrchyan:2008zzk} is a
superconducting solenoid of 6\unit{m} internal diameter, providing a
magnetic field of 3.8\unit{T}. A silicon pixel and strip tracker, a
lead tungstate crystal electromagnetic calorimeter (ECAL), and a brass
and scintillator hadron calorimeter, each composed of a barrel and two
endcap sections, are located within the solenoid volume. Muons are
measured in gas-ionization detectors embedded in the steel flux-return
yoke outside the solenoid. A two-tier trigger system selects the most
interesting \Pp\Pp\ collisions for offline analysis. A more detailed
description of the CMS detector, together with a definition of its
coordinate system and kinematic variables, can be found in
Ref.~\cite{Chatrchyan:2008zzk}.

We use several Monte Carlo (MC) generator programs to simulate signal
and background processes. The next-to-leading-order (NLO)
\POWHEG~(v2)~\cite{powheg,powheg2} generator is used to generate
\ttbar signal events, assuming a top quark mass of
$m_{\cPqt}=172.5\GeV$~\cite{worldave}. We utilize the
NNPDF3.0 NLO~\cite{nnpdf} parton distribution functions (PDF) in
the MC calculations. The events are interfaced to
\PYTHIA~(v8.205)~\cite{Sjostrand:2006za,Sjostrand:2014zea} with the
CUETP8M1 tune~\cite{CMS-PAS-GEN-14-001,Skands:2014pea} to simulate
parton showering, hadronization, and the underlying event. An
alternative sample is obtained using the
\HERWIGpp~(v2.7.1)~\cite{herwigpp} program to model the parton
shower. Another sample of \ttbar events is generated using
\amcatnlo~(v5\_2.2.2)~\cite{amcatnlo} and \madspin~\cite{madspin}
generators, and again \PYTHIA~(v8.205) for parton showering,
hadronization, and the underlying event. The MC generators have been
validated by comparing to unfolded differential distributions of
\ttbar production at $\sqrt{s}=8\TeV$~\cite{mcval}.

Background events are simulated by the \amcatnlo~(v5\_2.2.2) generator
for \PW+jets production and Drell--Yan (DY) quark-antiquark
annihilation into lepton-antilepton pairs through virtual photon or \Z
boson exchange, with normalization taken from data. Associated top
quark and \PW\ boson production (\cPqt\PW) is simulated using
\POWHEG~(v1)~\cite{powheg1,powheg3} and \PYTHIA~(v8.205), and is
normalized to the approximate next-to-next-to-leading-order (NNLO)
cross section~\cite{Kidonakis:2013zqa}. The contributions from \PW\PW,
\PW\cPZ\, and \cPZ\cPZ\ (referred to as \VV) processes are simulated
with \PYTHIA~(v8.205), and normalized to their NLO cross
sections~\cite{mcfm}. All other backgrounds are estimated from control
samples extracted from collision data. The simulated samples include
additional interactions per bunch crossing (pileup). On average, about
20 collisions per bunch crossing are present in our data.

The SM prediction for the \ttbar production cross section at
$\sqrt{s}=13\TeV$ is calculated with the \textsc{Top++}
program~\cite{top++} at NNLO in perturbative QCD, including soft-gluon
resummation at next-to-next-to-leading-log order
(NNLL)~\cite{ttxsec1,ttxsec2,ttxsec3,ttxsec4,ttxsec5,mitov}, assuming
$m_{\cPqt}= 172.5\GeV$. The result is $\sigma_{\ttbar}^\mathrm{NNLO+NNLL} = 832^{+20}_{-29}\,\text{(scale)}\pm 35\,\text{(PDF}+\alpha_s)\unit{pb}$. The expected yields
for signal in all figures and tables are normalized to this value. The
first uncertainty reflects uncertainties in the factorization and
renormalization scales, $\mu_{\rm F}$ and $\mu_{\rm R}$. The second
uncertainty, associated with the PDFs and strong coupling constant
$\alpha_{\rm s}$, is obtained by following the PDF4LHC
prescription~\cite{pdf4lhcInterim,pdf4lhcReport} using the MSTW2008 68\% CL
NNLO~\cite{Martin:2009iq,mstw08}, CT10 NNLO~\cite{Lai:2010vv,pdfsets},
and NNPDF2.3 5f FFN~\cite{Ball:2012cx} PDF sets.

At the trigger level, events are required to contain one electron and one
muon, where the electron has transverse momentum $\pt > 12\GeV$ and
the muon has $\pt > 17\GeV$, or the electron has $\pt > 17\GeV$ and
the muon has $\pt > 8\GeV$. Offline, particle candidates are
reconstructed with the CMS particle-flow (PF)
algorithm~\cite{PFPAS1,PFPAS2}. The PF algorithm reconstructs and 
identifies each individual particle using an optimized combination of
information from the various elements of the CMS detector. 

Events are selected to contain one electron~\cite{emid} and one
muon~\cite{muid} of opposite charge, both of which are required to
have $\pt > 20\GeV$ and $\abs{\eta} < 2.4$ (but excluding electrons
within a small region of $\abs{\eta}$ between the barrel and endcap
sections of the ECAL). The electron and muon candidates are required
to be sufficiently isolated from nearby jet activity as follows. For
each electron and muon candidate, a cone of $\Delta R = 0.3$ and
$\Delta R = 0.4$, respectively, is constructed around the direction of
the track at the event vertex, where $\Delta R$ is defined as
$\sqrt{\smash[b]{(\Delta \eta)^2 + (\Delta \phi)^2}}$, and $\Delta
\eta$ and $\Delta \phi$ are the distances in pseudorapidity and
azimuthal angle. Excluding the contribution from the lepton candidate,
the scalar sum of the \pt of all particle candidates that are inside 
$\Delta R$ and are consistent with arising from the chosen primary event vertex is
calculated to define a relative isolation discriminant,
$I_\text{rel}$, through the ratio of this sum to the $\pt$ of the
lepton candidate. The neutral-particle contribution to $I_\text{rel}$
is corrected for pileup based on the average energy density deposited
by neutral particles in the event. This corresponds to an average
$\pt$ from pileup determined event-by-event that is subtracted from
the summed scalar $\pt$ in the isolation cone. An electron and muon
candidate is selected if they have respective values of
$I_\text{rel}<0.11$ and $I_\text{rel}<0.12$.

In events with more than one pair of leptons passing the above
selection, the two  leptons of opposite charge and different flavor
with the largest \pt are selected for further study.  Events with
\Pgt\ leptons contribute to the measurement only if they decay to
electrons or muons that satisfy the selection requirements, and are
included in the MC simulations.

The efficiency of the lepton selection is measured using a
``tag-and-probe'' method in same-flavor dilepton events enriched in \Z
boson candidates, as described in Refs.~\cite{inclusWZ3pb,CMStt8}. 
Differences in the event topology with respect to \ttbar\ production 
are accounted for as a systematic uncertainty. 
In the current data set, the measured values for the combined
identification and isolation efficiencies are typically 92\% for
muons and 77\% for electrons. Based on a comparison of lepton
selection efficiencies in data and simulation, the event yield in
simulation is corrected using $\pt$- and $\eta$-dependent
data-to-simulation scale factors (SF) to provide consistency with
data. They have average values of 1.00 for muons and 0.96 for
electrons.

Candidate events with dilepton invariant masses of $m_{\Pe\Pgm} <
20\GeV$ are removed to suppress backgrounds, mainly from low-mass DY 
processes. Jets are reconstructed from the PF particle candidates using the  
anti-\kt clustering algorithm~\cite{antikt} with a distance parameter of 
0.4, optimized for the running conditions at higher center-of-mass
energy. The jet energy is corrected for pileup in a manner similar to
that used to find the energy within the lepton isolation cone. Jet
energy corrections are also applied as a function of jet $\pt$ and
$\eta$~\cite{JESPUB} to data and simulation. Events are required to
have at least two reconstructed jets with $\pt> 30\GeV$ and
$\abs{\eta}< 2.4$.

Backgrounds in this analysis arise primarily from \cPqt\PW, DY, and \VV
events in which at least two leptons are produced. Background yields
from \cPqt\PW\ and \VV events are estimated from simulation.  The
\empm DY background normalization is estimated from data using the
``$R_\text{out/in}$'' method~\cite{CMStt8,CMStt12,CMStt13}, where
events with \eepm and \mmpm final states are explored as follows. A
data-to-simulation normalization factor is estimated from the number
of events within the \Z boson mass window in data, and extrapolated to
the number of events outside the \Z mass window with corrections based
on control regions in data enriched in DY events. This factor is found
to be $1.04 \pm 0.16\stat$.

Other background sources, such as \ttbar or \PW+jets events with
decays into one lepton and jets, can contaminate the signal sample if
a jet is incorrectly reconstructed as a lepton, or an event contains a
lepton from the decay of bottom or charm hadrons. These are grouped
into the nonprompt-lepton category, together with contributions that
can arise, for example, from the  decays of mesons, photon conversions
to $\Pep\Pem$  pairs in the material of the detector, or effects
from detector resolution. The nonprompt-lepton background is estimated
from an extrapolation of a control region of same-sign (SS) dilepton
events to the signal region of opposite-sign (OS) dileptons. The SS
control region is defined using the same criteria as used for the
nominal signal region, except requiring $\Pe\Pgm$ pairs of the same
charge. The SS dilepton events predominantly contain at least one
misidentified lepton.  Other SM processes, such as DY, \cPqt\PW, \VV
and \ttbar dilepton production have significantly smaller
contributions, and are estimated using simulation. The scaling from
the SS control region in data to the signal region is 
performed using an extrapolation factor, extracted from MC simulation,
given by the ratio of the number of OS events with misidentified
leptons to the number of SS events with misidentified leptons. From
the eight same-sign events observed in data, the expected
contamination of $1.7 \pm 0.4$ events due to DY, \cPqt\PW, \VV and
\ttbar dilepton production is subtracted, and the result is
multiplied by the OS to SS ratio of $1.4 \pm 0.3$ to obtain an
estimate of $8.5 \pm 4.4$ nonprompt lepton events contaminating the
signal, including statistical and systematic uncertainties. This
agrees with predictions from MC simulations of semileptonic \ttbar
and \PW+jets events.

\begin{figure}[htbp!]
\centering
\includegraphics[width=0.49\textwidth]{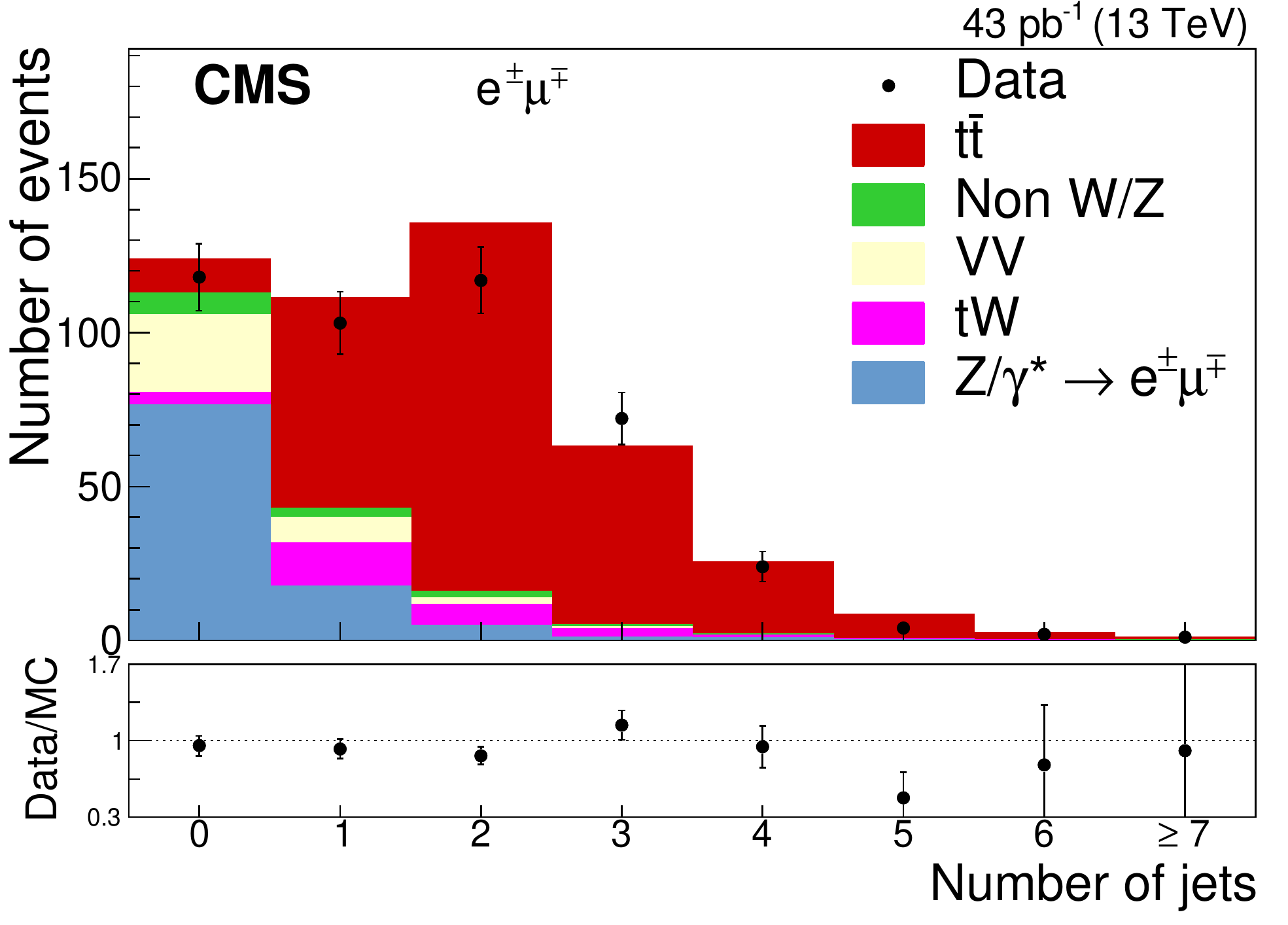}
\includegraphics[width=0.49\textwidth]{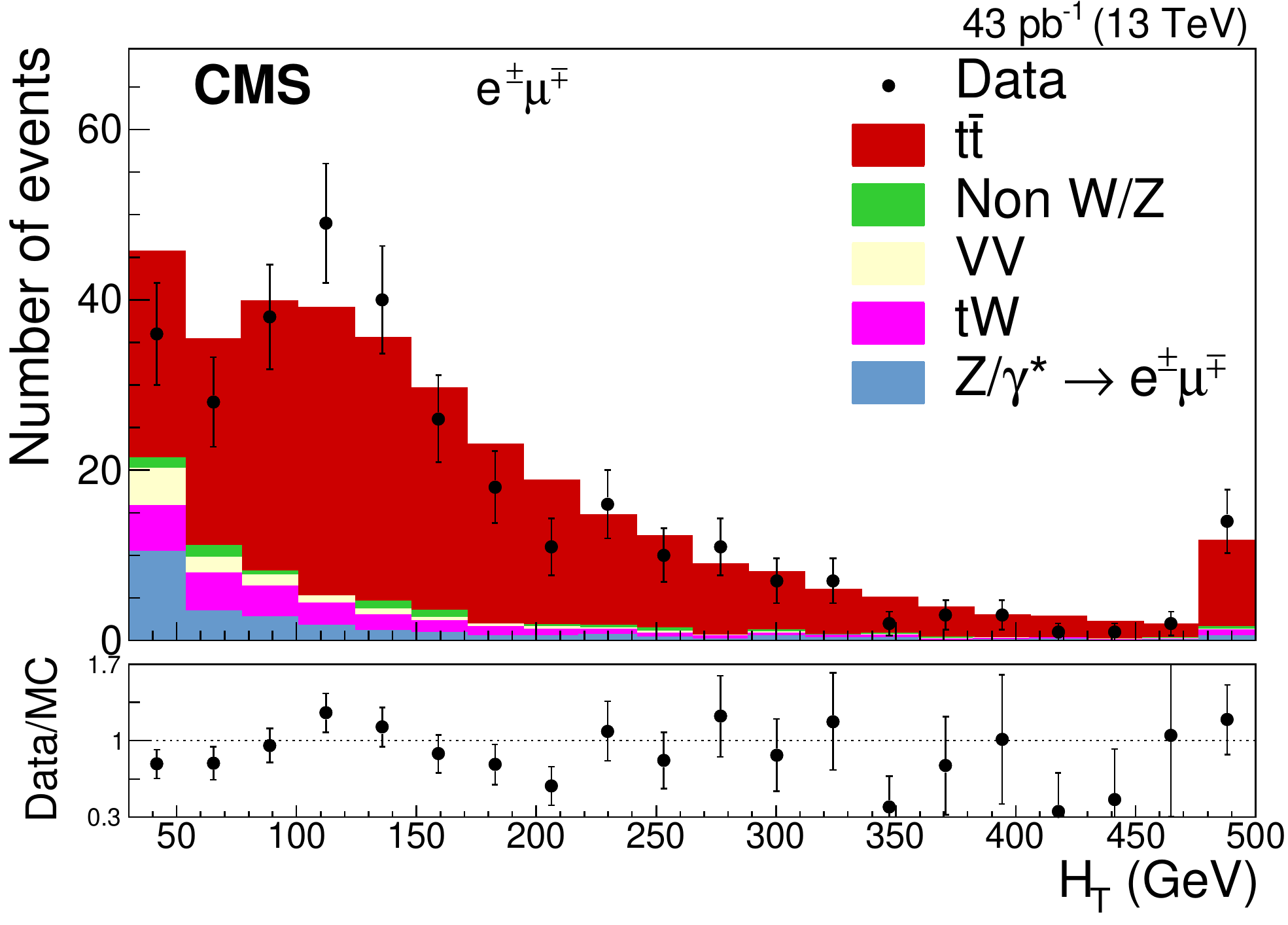}
\caption{The distributions in (\cmsLeft) the jet multiplicity, and (\cmsRight)
$H_\mathrm{T}$ in events passing the dilepton criteria. The expected
distributions for \ttbar signal and individual backgrounds are shown
after implementing data-based corrections; the last bin contains the
overflow in events. The ratios of data to the sum of the expected
yields are given at the bottom of each panel.}
\label{fig:dilepton}
\end{figure}

Figure~\ref{fig:dilepton} shows (\cmsLeft) the multiplicity of jets
and (\cmsRight) the scalar \pt sum of all jets ($H_{\rm T}$) for
events passing the dilepton criteria. Agreement is observed between
data and the predictions for signal and background.

\begin{figure}[htbp]
\centering
\includegraphics[width=0.49\textwidth]{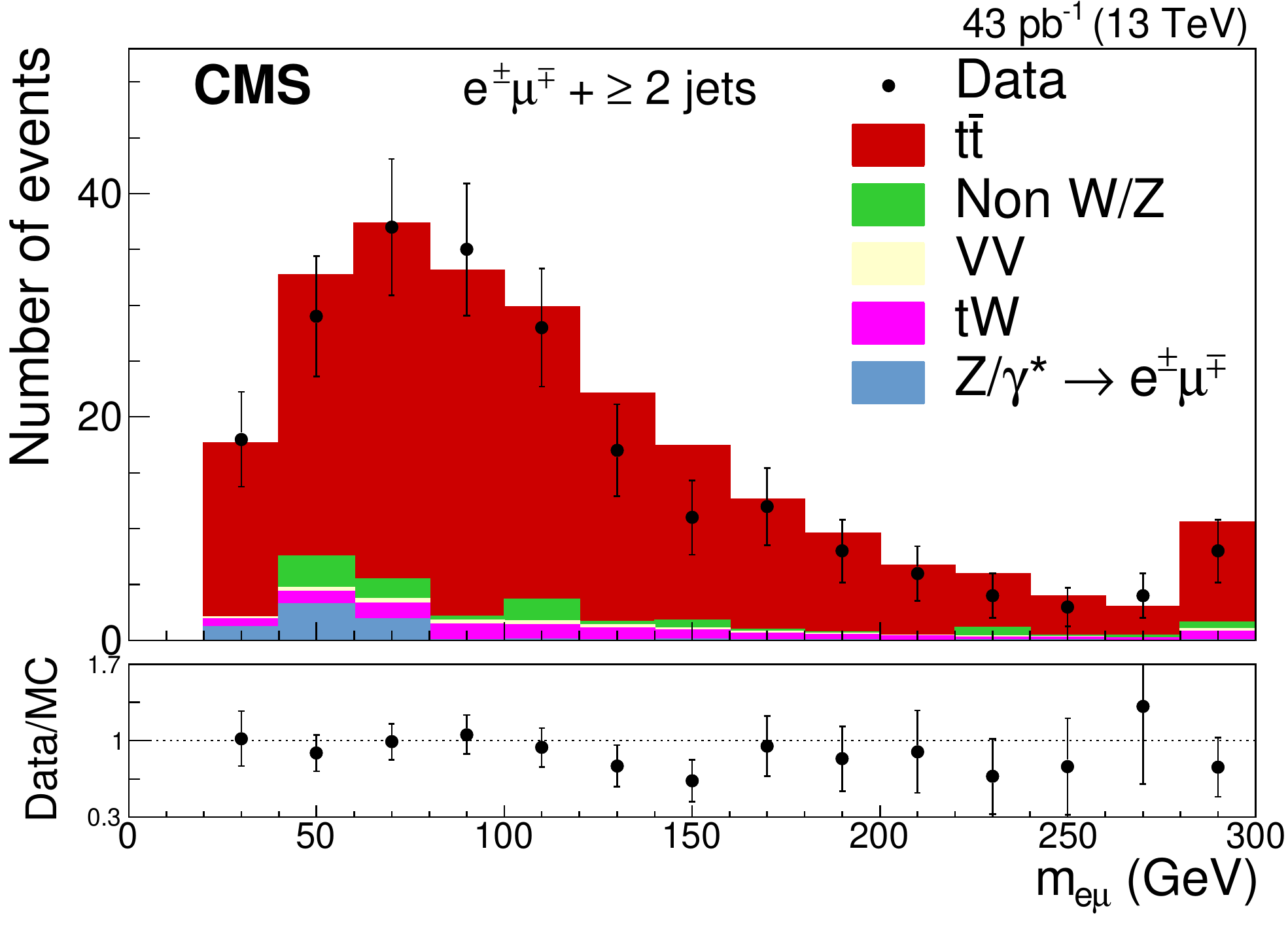}
\includegraphics[width=0.49\textwidth]{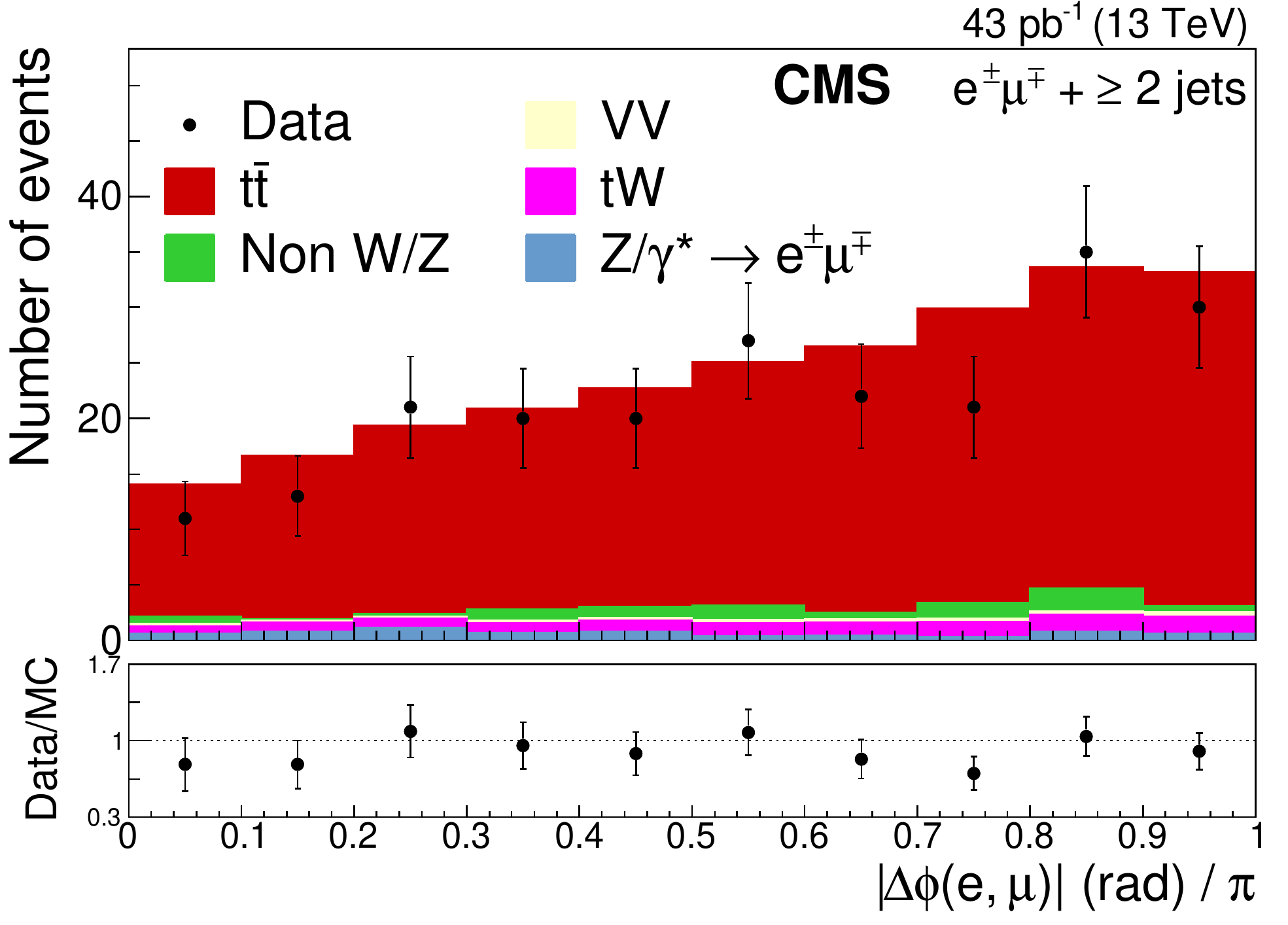}
\caption{The distributions in (\cmsLeft) the dilepton invariant mass, and (\cmsRight)
the difference in the azimuthal angle between the two leptons after
all selections. The last bin in (\cmsLeft) contains the overflow events. The
ratios of data to the sum of the expected yields are given at the
bottom of each panel.}
\label{fig:measured}
\end{figure}

After requiring at least two jets, we obtain the plots presented in
Fig.~\ref{fig:measured}, where (\cmsLeft) shows the distribution in the
invariant dilepton mass $m_{\Pe\Pgm}$, which is sensitive to the
existence of a new heavy object decaying into a \ttbar
pair. Figure~\ref{fig:measured} (\cmsRight) shows the difference in azimuthal
angle between the two leptons, $\Delta\phi(\Pe,\Pgm)$,
and explores the correlation between the \cPqt\ and \cPaqt\
spins~\cite{ttspin_parke,ttspin_bernreuther,ATLASnp3,ATLASspin7,ATLASspinobs,CMSspin7}.
For both distributions, data are in agreement with the SM
expectations.

The dominant uncertainty is due to the preliminary integrated
luminosity, which is estimated from $x$-$y$ beam-beam scans performed
in July 2015 utilizing the methods of Ref.~\cite{lumiPAS13}. The
resulting uncertainty in the integrated luminosity is 4.8\%.

Smaller uncertainties arise from the measured trigger efficiency, and
the lepton identification and isolation efficiencies. After the
offline dilepton selection, the trigger efficiency is measured in data
to be $(91\pm 4)\%$ using triggers based on the \pt imbalance in the
event.  This efficiency is applied to the MC simulations and the uncertainty is
taken as a global uncertainty. The uncertainties on the electron and
muon identification and  isolation efficiencies are estimated by
changing the $\pt$- and $\eta$-dependent SF values by one
standard deviation ($\pm 1 \sigma$). The modeling of lepton energy scales is studied
using $\Z \to\Pe\Pe$ and $\mu\mu$ events in data and
in simulation, yielding an uncertainty in the electron energy scale of
1\%, and in the muon energy scale of 0.5\%. The impact of the
uncertainty in the jet energy scale (JES) is estimated by changing the
\pt- and $\eta$-dependent JES SF by $\pm 1 \sigma$, and the uncertainty in jet energy resolution (JER)
uncertainty is estimated through similar $\eta$-dependent $\pm 1 \sigma$
changes in the JER SF. The maximum of each of the deviations is taken
as the uncertainty.

The distribution of the number of vertices per beam crossing is
compared between data and simulation. The results indicate agreement
of the total \Pp\Pp\ inelastic cross section within 10\%. The
result of varying this cross section by $\pm$10\% for all MC samples
is used to obtain the systematic uncertainty due to pileup.

Theory uncertainties on \ttbar production involve the systematic bias
related to the missing higher-order diagrams in \POWHEG, and is
estimated through studies of the signal acceptance by changing the
renormalization and factorization scales in \POWHEG 
simultaneously within the range $[ \mu/2 , 2\mu ]$ ($\mu = \mu_\mathrm{R}= \mu_\mathrm{F}$). 
In addition, the predictions of the NLO
generators \amcatnlo (v5\_2.2.2) and \POWHEG  are compared for
\ttbar production, where both use \PYTHIA~(v8.205) for hadronization,
parton showering, and simulation of the underlying event. The
uncertainty arising from the hadronization model mainly affects the
JES and the fragmentation of jets. The uncertainty in the JES already
contains a contribution from the uncertainty in the hadronization. The
hadronization uncertainty is also determined by comparing samples of
events generated with \POWHEG, where the hadronization is either
modeled with \PYTHIA~(v8.205) or \HERWIGpp~(v2.7.1). This also
includes differences in parton showering, and the underlying event,
and is called \ttbar modeling uncertainty. All theory uncertainties
on \ttbar production are taken as the maximum difference found in the
results. The uncertainty from the choice of PDF is determined by
reweighting the sample of simulated  \ttbar events according to the
26 CT10 NLO~\cite{Lai:2010vv,pdfsets} and the 100 NNPDF3.0
sets~\cite{nnpdf} of PDF uncertainties.

An uncertainty of 30\% in cross sections for \cPqt\PW\ and \VV
backgrounds are taken from measurements~\cite{CMSWWZZPublication8,
CMSWWPublication7,CMSWWHiggsPublication7,CMSWWZZPublication7,
CMSZZPublication7,
CMStopPublicationstop4,ATLASWWPublication,ATLASWZPublication,ATLASZZPublication}.
For DY production, a global cross section uncertainty of 15\% is
applied, which is derived from  the variation of the SF for events
passing the dilepton criteria and events passing all selection
cuts. The systematic uncertainty in the estimated nonprompt lepton
background is given mainly by the systematic uncertainty in the ratio
of OS to SS events with misidentified leptons in the MC
simulations. We checked how well the simulation models the production
of misidentified leptons by examining additional control regions, with
the observed discrepancy used to assign an uncertainty of 23\% to the
method.

Table~\ref{tab:breakdown_comb}  summarizes the magnitude of the
statistical and systematic uncertainties from different sources
contributing to the \ttbar production cross section. All sources of
uncertainties are added in quadrature.

\begin{table}[htbp!]
\centering
\topcaption{Summary of individual contributions to the systematic
uncertainty in the $\sigma_{\ttbar}$  measurement. The uncertainties
are given in pb, and as relative uncertainties. The separate total
systematic uncertainty without integrated luminosity, the part
attributed to the integrated luminosity, and the statistical
contributions are added in quadrature to obtain the total uncertainty.}
\begin{scotch}{lrr}
Source                           &  $\Delta\sigma_{\ttbar}$ (pb)  & \multicolumn{1}{c}{$\Delta\sigma_{\ttbar} /\sigma_{\ttbar}$ (\%)} \\
\hline
Trigger efficiencies             &  33  &  4.4 \\
Lepton efficiencies              &  25  &  3.4 \\
Lepton energy scale              & $<$1 & $\le$0.1 \\
Jet energy scale                 &  11  &  1.5 \\
Jet energy resolution            & $<$1 & $\le$0.1 \\
Pileup                           &  5.2 &  0.7 \\
QCD scales                       &  1.4 &  0.2 \\
NLO generator of \ttbar signal  &  14  &  1.9 \\
Modeling of \ttbar signal       &  13  &  1.8 \\
PDF                              &  18  &  2.4 \\
Single top \cPqt\PW\ background  &  13  &  1.8 \\
\VV background                    &  3.5 &  0.5 \\
Drell-Yan background             &  4.1 &  0.5 \\
Nonprompt leptons background     &  7.6 &  1.0 \\
\hline
Total systematic                 & \multirow{2}{*}{ 53 } & \multirow{2}{*}{ 7.2 }\\
(w/o luminosity)                 &      &  \\
\hline
Integrated luminosity            & 36   &   4.8 \\
\hline
Statistical uncertainty          & 58   &  7.8 \\
\hline
Total                            & 87  &  \relerr \\
\end{scotch}
\label{tab:breakdown_comb}
\end{table}

Table~\ref{tab:yields} shows the total number of events observed in
data, together with the total number of background events expected
from simulation or estimated from data. The mean acceptance multiplied
by the selection efficiency and the branching fraction, as estimated
from simulation at $m_{\PQt} = 172.5$\GeV, is $\epsilon = (0.60\pm 0.04)\%$, including
statistical and systematic uncertainties. The measured fiducial cross
section for \ttbar production with two leptons (one electron and one
muon) in the range $\pt > 20\GeV$ and $\abs{\eta} < 2.4$, is
$\sigma^\text{fid}_{\ttbar} = 12.4 \pm 1.0\stat \pm 1.0\syst\pm
0.6\lum\unit{pb}$. After applying all corrections, the inclusive cross
section is measured to be \xsec.

\begin{table}[!htb]
\centering
\topcaption{The number of \Pe\Pgm\ events after final event selection
expected for background, and observed in data. The uncertainties
represent the statistical and systematic components added in
quadrature.}
\newcolumntype{x}{D{,}{\,\pm\,}{-1}}
\begin{scotch}{lx}
Source &  \multicolumn{1}{c}{Number of events}  \\
       & \multicolumn{1}{c}{\empm}            \\
\hline
Drell--Yan          &   6.9,1.2 \\
Nonprompt leptons  &   8.5,4.4 \\
\cPqt\PW\          &  10.9,3.4 \\
\VV                 &   2.7,0.9\\
\hline
Total background   &  29.1,5.7\\
\hline
Data               & \multicolumn{1}{c}{220}                \\
\end{scotch}
\label{tab:yields}
\end{table}

A linear parametrization of the acceptance dependence on $m_{\PQt}$
in the range 169.5--175.5\GeV results in a cross section reduction of
${\approx}0.7\%$ at $m_{\PQt} = 173.34$\GeV, the current world
average of the top quark mass~\cite{worldave}.

In an alternative analysis, the selected sample is split into events
with 0, 1, 2, and $>2$ \PQb quark jets, and 0, 1, 2, and $>$2 additional
light-flavor or gluon jets (\ie, not identified as b quark jets). Jets
are identified as b quark jets using the combined secondary vertex
(CSV) algorithm~\cite{BTV-11-004-pub}. A maximum likelihood fit of the
yields in different input samples is performed to extract
simultaneously $\sigma_{\ttbar}$ and the b tagging
efficiency. Systematic uncertainties are implemented through nuisance
parameters~\cite{nuisance}. This result is within 1\% of the nominal
analysis.

Figure~1 in \suppMaterial
presents a summary of results for $\sigma_{\ttbar}$ from the
combination of the Tevatron measurements at
$1.96\TeV$~\cite{tevatron_combi}, from CMS measurements at $\sqrt{s} =
7$ and 8\TeV~\cite{CMStt3,CMStt8}, and from the measurement
presented here at $\sqrt{s} = 13\TeV$, compared to the NNLO+NNLL
predictions as a function of $\sqrt{s}$ for $\Pp\Pap$ and $\Pp\Pp$ collisions~\cite{mitov}.

In summary, the first measurement of the \ttbar production cross
section in proton-proton collisions at $\sqrt{s} =13\TeV$ is presented
for events containing an electron-muon pair and at least two jets. The
measurement is obtained through an event-counting analysis based on a
data sample corresponding to an integrated luminosity of
\usedLumi. The result is \xsec, with a total relative uncertainty of
\relerr\%. This measurement is consistent with the SM prediction of
$\sigma_{\ttbar}^\mathrm{NNLO+NNLL} = 832^{+40}_{-46}\unit{pb}$ for a top quark mass of 172.5\GeV.

We congratulate our colleagues in the CERN accelerator departments for the excellent performance of the LHC and thank the technical and administrative staffs at CERN and at other CMS institutes for their contributions to the success of the CMS effort. In addition, we gratefully acknowledge the computing centers and personnel of the Worldwide LHC Computing Grid for delivering so effectively the computing infrastructure essential to our analyses. Finally, we acknowledge the enduring support for the construction and operation of the LHC and the CMS detector provided by the following funding agencies: BMWFW and FWF (Austria); FNRS and FWO (Belgium); CNPq, CAPES, FAPERJ, and FAPESP (Brazil); MES (Bulgaria); CERN; CAS, MoST, and NSFC (China); COLCIENCIAS (Colombia); MSES and CSF (Croatia); RPF (Cyprus); MoER, ERC IUT and ERDF (Estonia); Academy of Finland, MEC, and HIP (Finland); CEA and CNRS/IN2P3 (France); BMBF, DFG, and HGF (Germany); GSRT (Greece); OTKA and NIH (Hungary); DAE and DST (India); IPM (Iran); SFI (Ireland); INFN (Italy); MSIP and NRF (Republic of Korea); LAS (Lithuania); MOE and UM (Malaysia); CINVESTAV, CONACYT, SEP, and UASLP-FAI (Mexico); MBIE (New Zealand); PAEC (Pakistan); MSHE and NSC (Poland); FCT (Portugal); JINR (Dubna); MON, RosAtom, RAS and RFBR (Russia); MESTD (Serbia); SEIDI and CPAN (Spain); Swiss Funding Agencies (Switzerland); MST (Taipei); ThEPCenter, IPST, STAR and NSTDA (Thailand); TUBITAK and TAEK (Turkey); NASU and SFFR (Ukraine); STFC (United Kingdom); DOE and NSF (USA).

\bibliography{auto_generated}

\providecommand{\href}[2]{#2}\begingroup\raggedright\begin{thebibliography}{10}%
\makeatletter
\providecommand{\hrefCMSnoop }[0]{\@secondoftwo}%
\makeatother
\providecommand{\doi}{\texttt{doi:}\begingroup \urlstyle{tt}\Url}

\bibitem{ATLAStt1}
\hrefCMSnoop {}{{ATLAS Collaboration}, ``{Measurement of the top pair
  production cross section in 8 TeV proton-proton collisions using kinematic
  information in the lepton+jets final state with ATLAS}'',} \textit{ Phys.
  Rev. D} \textbf{ 91} (2015) 112013,
  \href{http://dx.doi.org/10.1103/PhysRevD.91.112013}{\doi{10.1103/PhysRevD.91.112013}},
\href{http://www.arXiv.org/abs/1504.04251}{\texttt{arXiv:1504.04251}}.

\bibitem{ATLAStt2}
\hrefCMSnoop {}{{ATLAS Collaboration}, ``{Differential top-antitop
  cross-section measurements as a function of observables constructed from
  final-state particles using pp collisions at $\sqrt{s}= 7\TeV$ in the ATLAS
  detector}'',} \textit{ JHEP} \textbf{ 06} (2015) 100,
  \href{http://dx.doi.org/10.1007/JHEP06(2015)100}{\doi{10.1007/JHEP06(2015)100}},
\href{http://www.arXiv.org/abs/1502.05923}{\texttt{arXiv:1502.05923}}.

\bibitem{ATLAStt3}
\hrefCMSnoop {}{{ATLAS Collaboration}, ``{Measurement of the $\rm t\overline{t}
  $ production cross-section as a function of jet multiplicity and jet
  transverse momentum in 7 TeV proton-proton collisions with the ATLAS
  detector}'',} \textit{ JHEP} \textbf{ 01} (2015) 020,
  \href{http://dx.doi.org/10.1007/JHEP01(2015)020}{\doi{10.1007/JHEP01(2015)020}},
\href{http://www.arXiv.org/abs/1407.0891}{\texttt{arXiv:1407.0891}}.

\bibitem{ATLAStt4}
\hrefCMSnoop {}{{ATLAS Collaboration}, ``{Simultaneous measurements of the $\rm
  t\bar{t}$, $\rm W^+W^-$, and $\cPZ / \cPgg^{*} \rightarrow \PGt\PGt$
  production cross-sections in $\rm pp$ collisions at $\sqrt{s}= 7\TeV$ with
  the ATLAS detector}'',} \textit{ Phys. Rev. D} \textbf{ 91} (2015) 052005,
  \href{http://dx.doi.org/10.1103/PhysRevD.91.052005}{\doi{10.1103/PhysRevD.91.052005}},
\href{http://www.arXiv.org/abs/1407.0573}{\texttt{arXiv:1407.0573}}.

\bibitem{ATLAStt5}
\hrefCMSnoop {}{{ATLAS Collaboration}, ``{Measurements of normalized
  differential cross sections for $\rm t\bar{t}$ production in pp collisions at
  $\sqrt{s}= 7\TeV$ using the ATLAS detector}'',} \textit{ Phys. Rev. D}
  \textbf{ 90} (2014) 072004,
  \href{http://dx.doi.org/10.1103/PhysRevD.90.072004}{\doi{10.1103/PhysRevD.90.072004}},
\href{http://www.arXiv.org/abs/1407.0371}{\texttt{arXiv:1407.0371}}.

\bibitem{ATLAStt6}
\hrefCMSnoop {}{{ATLAS Collaboration}, ``{Measurement of the $\rm
  t\overline{t}$ production cross-section using \Pe\Pgm\ events with $\rm
  b$-tagged jets in $\rm pp$ collisions at $\sqrt{s}=7$ and $8\TeV$ with the
  ATLAS detector}'',} \textit{ Eur. Phys. J. C} \textbf{ 74} (2014) 3109,
  \href{http://dx.doi.org/10.1140/epjc/s10052-014-3109-7}{\doi{10.1140/epjc/s10052-014-3109-7}},
\href{http://www.arXiv.org/abs/1406.5375}{\texttt{arXiv:1406.5375}}.

\bibitem{ATLAStt8}
\hrefCMSnoop {}{{ATLAS Collaboration}, ``{Measurement of the \ttbar production
  cross section in the tau+jets channel using the ATLAS detector}'',} \textit{
  Eur. Phys. J. C} \textbf{ 73} (2013) 2328,
  \href{http://dx.doi.org/10.1140/epjc/s10052-013-2328-7}{\doi{10.1140/epjc/s10052-013-2328-7}},
\href{http://www.arXiv.org/abs/1211.7205}{\texttt{arXiv:1211.7205}}.

\bibitem{ATLAStt9}
\hrefCMSnoop {}{{ATLAS Collaboration}, ``{Measurement of the top quark pair
  cross section with ATLAS in pp collisions at $\sqrt{s} = 7\TeV$ using final
  states with an electron or a muon and a hadronically decaying \PGt\
  lepton}'',} \textit{ Phys. Lett. B} \textbf{ 717} (2012) 89,
  \href{http://dx.doi.org/10.1016/j.physletb.2012.09.032}{\doi{10.1016/j.physletb.2012.09.032}},
\href{http://www.arXiv.org/abs/1205.2067}{\texttt{arXiv:1205.2067}}.

\bibitem{ATLAStt10}
\hrefCMSnoop {}{{ATLAS Collaboration}, ``{Measurement of $\rm t \bar{t}$
  production with a veto on additional central jet activity in pp collisions at
  $\sqrt{s}= 7\TeV$ using the ATLAS detector}'',} \textit{ Eur. Phys. J. C}
  \textbf{ 72} (2012) 2043,
  \href{http://dx.doi.org/10.1140/epjc/s10052-012-2043-9}{\doi{10.1140/epjc/s10052-012-2043-9}},
\href{http://www.arXiv.org/abs/1203.5015}{\texttt{arXiv:1203.5015}}.

\bibitem{ATLAStt11}
\hrefCMSnoop {}{{ATLAS Collaboration}, ``{Measurement of the cross section for
  top-quark pair production in $\rm pp$ collisions at $\sqrt{s}= 7\TeV$ with
  the ATLAS detector using final states with two high-\pt leptons}'',} \textit{
  JHEP} \textbf{ 05} (2012) 059,
  \href{http://dx.doi.org/10.1007/JHEP05(2012)059}{\doi{10.1007/JHEP05(2012)059}},
\href{http://www.arXiv.org/abs/1202.4892}{\texttt{arXiv:1202.4892}}.

\bibitem{ATLAStt12}
\hrefCMSnoop {}{{ATLAS Collaboration}, ``{Measurement of the top quark pair
  production cross-section with ATLAS in the single lepton channel}'',}
  \textit{ Phys. Lett. B} \textbf{ 711} (2012) 244,
  \href{http://dx.doi.org/10.1016/j.physletb.2012.03.083}{\doi{10.1016/j.physletb.2012.03.083}},
\href{http://www.arXiv.org/abs/1201.1889}{\texttt{arXiv:1201.1889}}.

\bibitem{CMStt1}
\hrefCMSnoop {}{{CMS Collaboration}, ``{Measurement of the differential cross
  section for top quark pair production in pp collisions at $\sqrt{s}=
  8\TeV$}'',} (2015).
\href{http://www.arXiv.org/abs/1505.04480}{\texttt{arXiv:1505.04480}}.

\bibitem{CMStt2}
\hrefCMSnoop {}{{CMS Collaboration}, ``{Measurement of the $\rm t \bar t$
  production cross section in $\rm pp$ collisions at $\sqrt{s} = 8\TeV$ in
  dilepton final states containing one \PGt\ lepton}'',} \textit{ Phys. Lett.
  B} \textbf{ 739} (2014) 23,
  \href{http://dx.doi.org/10.1016/j.physletb.2014.10.032}{\doi{10.1016/j.physletb.2014.10.032}},
\href{http://www.arXiv.org/abs/1407.6643}{\texttt{arXiv:1407.6643}}.

\bibitem{CMStt3}
\hrefCMSnoop {}{{CMS Collaboration}, ``{Measurement of the $\rm t \bar{t}$
  production cross section in the dilepton channel in pp collisions at
  $\sqrt{s} = 8\TeV$}'',} \textit{ JHEP} \textbf{ 02} (2014) 024,
  \href{http://dx.doi.org/10.1007/JHEP02(2014)024}{\doi{10.1007/JHEP02(2014)024}},
  \href{http://www.arXiv.org/abs/1312.7582}{\texttt{arXiv:1312.7582}}.
[Erratum: \DOI{10.1007/JHEP02(2014)102}].

\bibitem{CMStt4}
\hrefCMSnoop {}{{CMS Collaboration}, ``{Measurement of the $\rm t\bar{t}$
  production cross section in the all-jet final state in pp collisions at
  $\sqrt{s} = 7 \TeV$}'',} \textit{ JHEP} \textbf{ 05} (2013) 065,
  \href{http://dx.doi.org/10.1007/JHEP05(2013)065}{\doi{10.1007/JHEP05(2013)065}},
\href{http://www.arXiv.org/abs/1302.0508}{\texttt{arXiv:1302.0508}}.

\bibitem{CMStt5}
\hrefCMSnoop {}{{CMS Collaboration}, ``{Measurement of the \ttbar\ production
  cross section in the \PGt+jets channel in pp collisions at $\sqrt{s} =
  7\TeV$}'',} \textit{ Eur. Phys. J. C} \textbf{ 73} (2013) 2386,
  \href{http://dx.doi.org/10.1140/epjc/s10052-013-2386-x}{\doi{10.1140/epjc/s10052-013-2386-x}},
\href{http://www.arXiv.org/abs/1301.5755}{\texttt{arXiv:1301.5755}}.

\bibitem{CMStt6}
\hrefCMSnoop {}{{CMS Collaboration}, ``{Measurement of the $\rm t\bar{t}$
  production cross section in $\rm pp$ collisions at $\sqrt{s}= 7\TeV$ with
  lepton + jets final states}'',} \textit{ Phys. Lett. B} \textbf{ 720} (2013)
  83,
  \href{http://dx.doi.org/10.1016/j.physletb.2013.02.021}{\doi{10.1016/j.physletb.2013.02.021}},
\href{http://www.arXiv.org/abs/1212.6682}{\texttt{arXiv:1212.6682}}.

\bibitem{CMStt7}
\hrefCMSnoop {}{{CMS Collaboration}, ``{Measurement of differential
  top-quark-pair production cross sections in $\rm pp$ colisions at $\sqrt{s}=
  7\TeV$}'',} \textit{ Eur. Phys. J. C} \textbf{ 73} (2013) 2339,
  \href{http://dx.doi.org/10.1140/epjc/s10052-013-2339-4}{\doi{10.1140/epjc/s10052-013-2339-4}},
\href{http://www.arXiv.org/abs/1211.2220}{\texttt{arXiv:1211.2220}}.

\bibitem{CMStt8}
\hrefCMSnoop {}{{CMS Collaboration}, ``{Measurement of the $\rm t\bar{t}$
  production cross section in the dilepton channel in $\rm pp$ collisions at
  $\sqrt{s}=7\TeV$}'',} \textit{ JHEP} \textbf{ 11} (2012) 067,
  \href{http://dx.doi.org/10.1007/JHEP11(2012)067}{\doi{10.1007/JHEP11(2012)067}},
\href{http://www.arXiv.org/abs/1208.2671}{\texttt{arXiv:1208.2671}}.

\bibitem{CMStt9}
\hrefCMSnoop {}{{CMS Collaboration}, ``{Measurement of the \ttbar\ production
  cross section in $\rm pp$ collisions at $\sqrt{s}= 7\TeV$ in dilepton final
  states containing a \PGt}'',} \textit{ Phys. Rev. D} \textbf{ 85} (2012)
  112007,
  \href{http://dx.doi.org/10.1103/PhysRevD.85.112007}{\doi{10.1103/PhysRevD.85.112007}},
\href{http://www.arXiv.org/abs/1203.6810}{\texttt{arXiv:1203.6810}}.

\bibitem{Chatrchyan:2008zzk}
\hrefCMSnoop {}{{CMS Collaboration}, ``The {CMS} experiment at the {CERN}
  {LHC}'',} \textit{ JINST} \textbf{ 3} (2008) S08004,
  \href{http://dx.doi.org/10.1088/1748-0221/3/08/S08004}{\doi{10.1088/1748-0221/3/08/S08004}}.

\bibitem{powheg}
\hrefCMSnoop {}{S.~Frixione, P.~Nason, and C.~Oleari, ``{Matching NLO QCD
  computations with parton shower simulations: the POWHEG method}'',} \textit{
  JHEP} \textbf{ 11} (2007) 070,
  \href{http://dx.doi.org/10.1088/1126-6708/2007/11/070}{\doi{10.1088/1126-6708/2007/11/070}},
  \href{http://www.arXiv.org/abs/0709.2092}{\texttt{arXiv:0709.2092}}.

\bibitem{powheg2}
\hrefCMSnoop {}{S.~Alioli, P.~Nason, C.~Oleari, and E.~Re, ``{A general
  framework for implementing NLO calculations in shower Monte Carlo programs:
  the POWHEG BOX}'',} \textit{ JHEP} \textbf{ 06} (2010) 043,
  \href{http://dx.doi.org/10.1007/JHEP06(2010)043}{\doi{10.1007/JHEP06(2010)043}},
\href{http://www.arXiv.org/abs/1002.2581}{\texttt{arXiv:1002.2581}}.

\bibitem{worldave}
\hrefCMSnoop {}{{ATLAS, CDF, CMS and D0 collaborations}, ``{First combination
  of Tevatron and LHC measurements of the top-quark mass}'',} (2014).
\href{http://www.arXiv.org/abs/1403.4427}{\texttt{arXiv:1403.4427}}.

\bibitem{nnpdf}
F.~Demartin\hrefCMSnoop {}{ {et~al.}, ``{Impact of parton distribution function
  and $\alpha_S$ uncertainties on Higgs boson production in gluon fusion at
  hadron colliders}'',} \textit{ Phys. Rev. D} \textbf{ 82} (2010) 014002,
  \href{http://dx.doi.org/10.1103/PhysRevD.82.014002}{\doi{10.1103/PhysRevD.82.014002}},
\href{http://www.arXiv.org/abs/1004.0962}{\texttt{arXiv:1004.0962}}.

\bibitem{Sjostrand:2006za}
\hrefCMSnoop {}{T.~Sj{\"o}strand, S.~Mrenna, and P.~Skands, ``{PYTHIA 6.4
  physics and manual}'',} \textit{ JHEP} \textbf{ 05} (2006) 026,
  \href{http://dx.doi.org/10.1088/1126-6708/2006/05/026}{\doi{10.1088/1126-6708/2006/05/026}},
\href{http://www.arXiv.org/abs/hep-ph/0603175}{\texttt{arXiv:hep-ph/0603175}}.

\bibitem{Sjostrand:2014zea}
T.~Sj{\"o}strand\hrefCMSnoop {}{ {et~al.}, ``{An introduction to PYTHIA
  8.2}'',} \textit{ Comput. Phys. Commun.} \textbf{ 191} (2015) 159,
  \href{http://dx.doi.org/10.1016/j.cpc.2015.01.024}{\doi{10.1016/j.cpc.2015.01.024}},
\href{http://www.arXiv.org/abs/1410.3012}{\texttt{arXiv:1410.3012}}.

\bibitem{CMS-PAS-GEN-14-001}
\href {https://cds.cern.ch/record/1697700}{{CMS Collaboration}, ``{Underlying
  Event Tunes and Double Parton Scattering}'',} CMS Physics Analysis Summary
  CMS-PAS-GEN-14-001, 2014.

\bibitem{Skands:2014pea}
\hrefCMSnoop {}{P.~Skands, S.~Carrazza, and J.~Rojo, ``{Tuning PYTHIA 8.1: the
  Monash 2013 tune}'',} \textit{ Eur. Phys. J. C} \textbf{ 74} (2014) 3024,
  \href{http://dx.doi.org/10.1140/epjc/s10052-014-3024-y}{\doi{10.1140/epjc/s10052-014-3024-y}},
\href{http://www.arXiv.org/abs/1404.5630}{\texttt{arXiv:1404.5630}}.

\bibitem{herwigpp}
M.~B{\"a}hr\hrefCMSnoop {}{ {et~al.}, ``Herwig++ physics and manual'',}
  \textit{ Eur. Phys. J. C} \textbf{ 58} (2008) 639,
  \href{http://dx.doi.org/10.1140/epjc/s10052-008-0798-9}{\doi{10.1140/epjc/s10052-008-0798-9}},
\href{http://www.arXiv.org/abs/0803.0883}{\texttt{arXiv:0803.0883}}.

\bibitem{amcatnlo}
J.~Alwall\hrefCMSnoop {}{ {et~al.}, ``{The automated computation of tree-level
  and next-to-leading order differential cross sections, and their matching to
  parton shower simulations}'',} \textit{ JHEP} \textbf{ 07} (2014) 079,
  \href{http://dx.doi.org/10.1007/JHEP07(2014)079}{\doi{10.1007/JHEP07(2014)079}},
\href{http://www.arXiv.org/abs/1405.0301}{\texttt{arXiv:1405.0301}}.

\bibitem{madspin}
\hrefCMSnoop {}{P.~Artoisenet, R.~Frederix, O.~Mattelaer, and R.~Rietkerk,
  ``{Automatic spin-entangled decays of heavy resonances in Monte Carlo
  simulations}'',} \textit{ JHEP} \textbf{ 03} (2013) 015,
  \href{http://dx.doi.org/10.1007/JHEP03(2013)015}{\doi{10.1007/JHEP03(2013)015}},
  \href{http://www.arXiv.org/abs/1212.3460}{\texttt{arXiv:1212.3460}}.

\bibitem{mcval}
\hrefCMSnoop {}{{CMS Collaboration}, ``{Measurement of \ttbar\ production with
  additional jet activity, including b quark jets, in the dilepton channel
  using pp collisions at $\sqrt{s} = 8\TeV$}'',} (2015).
  \href{http://www.arXiv.org/abs/1510.03072}{\texttt{arXiv:1510.03072}}.
Submitted to Eur. Phys. J. C.

\bibitem{powheg1}
\hrefCMSnoop {}{S.~Alioli, P.~Nason, C.~Oleari, and E.~Re, ``{NLO single-top
  production matched with shower in POWHEG: $s$- and $t$-channel
  contributions}'',} \textit{ JHEP} \textbf{ 09} (2009) 111,
  \href{http://dx.doi.org/10.1088/1126-6708/2009/09/111}{\doi{10.1088/1126-6708/2009/09/111}},
  \href{http://www.arXiv.org/abs/0907.4076}{\texttt{arXiv:0907.4076}}.
[Erratum: \DOI{10.1007/JHEP02(2010)011}].

\bibitem{powheg3}
\hrefCMSnoop {}{E.~Re, ``{Single-top Wt-channel production matched with parton
  showers using the POWHEG method}'',} \textit{ Eur. Phys. J. C} \textbf{ 71}
  (2011) 1547,
  \href{http://dx.doi.org/10.1140/epjc/s10052-011-1547-z}{\doi{10.1140/epjc/s10052-011-1547-z}},
  \href{http://www.arXiv.org/abs/1009.2450}{\texttt{arXiv:1009.2450}}.

\bibitem{Kidonakis:2013zqa}
\hrefCMSnoop {}{N.~Kidonakis, ``{Top Quark Production}'',} in \textit{
  {Proceedings, Helmholtz International Summer School on Physics of Heavy
  Quarks and Hadrons (HQ 2013)}}, p.~139.
\newblock Verlag Deutsches Elektronen-Synchrotron, Hamburg, 2014.
\newblock \href{http://www.arXiv.org/abs/1311.0283}{\texttt{arXiv:1311.0283}}.
\newblock
\href{http://dx.doi.org/10.3204/DESY-PROC-2013-03/Kidonakis}{\doi{10.3204/DESY-PROC-2013-03/Kidonakis}}.

\bibitem{mcfm}
\hrefCMSnoop {}{J.~M. Campbell and R.~K. Ellis, ``{MCFM for the Tevatron and
  the LHC}'',} \textit{ Nucl. Phys. Proc. Suppl.} \textbf{ 205} (2010) 10,
  \href{http://dx.doi.org/10.1016/j.nuclphysbps.2010.08.011}{\doi{10.1016/j.nuclphysbps.2010.08.011}},
  \href{http://www.arXiv.org/abs/1007.3492}{\texttt{arXiv:1007.3492}}.

\bibitem{top++}
\hrefCMSnoop {}{M.~Czakon and A.~Mitov, ``{Top++: a program for the calculation
  of the top-pair cross-section at hadron colliders}'',} \textit{ Comput. Phys.
  Commun.} \textbf{ 185} (2014) 2930,
  \href{http://dx.doi.org/10.1016/j.cpc.2014.06.021}{\doi{10.1016/j.cpc.2014.06.021}},
\href{http://www.arXiv.org/abs/1112.5675}{\texttt{arXiv:1112.5675}}.

\bibitem{ttxsec1}
\hrefCMSnoop {}{M.~Beneke, P.~Falgari, S.~Klein, and C.~Schwinn, ``{Hadronic
  top-quark pair production with NNLL threshold resummation}'',} \textit{ Nucl.
  Phys. B} \textbf{ 855} (2012) 695,
  \href{http://dx.doi.org/10.1016/j.nuclphysb.2011.10.021}{\doi{10.1016/j.nuclphysb.2011.10.021}},
\href{http://www.arXiv.org/abs/1109.1536}{\texttt{arXiv:1109.1536}}.

\bibitem{ttxsec2}
M.~Cacciari\hrefCMSnoop {}{ {et~al.}, ``{Top-pair production at hadron
  colliders with next-to-next-to-leading logarithmic soft-gluon
  resummation}'',} \textit{ Phys. Lett. B} \textbf{ 710} (2012) 612,
  \href{http://dx.doi.org/10.1016/j.physletb.2012.03.013}{\doi{10.1016/j.physletb.2012.03.013}},
\href{http://www.arXiv.org/abs/1111.5869}{\texttt{arXiv:1111.5869}}.

\bibitem{ttxsec3}
\hrefCMSnoop {}{P.~Baernreuther, M.~Czakon, and A.~Mitov,
  ``{Percent-Level-Precision Physics at the Tevatron: Next-to-Next-to-Leading
  Order QCD Corrections to $\rm q \bar{q} \to t \bar{t} + X$}'',} \textit{
  Phys. Rev. Lett.} \textbf{ 109} (2012) 132001,
  \href{http://dx.doi.org/10.1103/PhysRevLett.109.132001}{\doi{10.1103/PhysRevLett.109.132001}},
\href{http://www.arXiv.org/abs/1204.5201}{\texttt{arXiv:1204.5201}}.

\bibitem{ttxsec4}
\hrefCMSnoop {}{M.~Czakon and A.~Mitov, ``{NNLO corrections to top-pair
  production at hadron colliders: the all-fermionic scattering channels}'',}
  \textit{ JHEP} \textbf{ 12} (2012) 054,
  \href{http://dx.doi.org/10.1007/JHEP12(2012)054}{\doi{10.1007/JHEP12(2012)054}},
\href{http://www.arXiv.org/abs/1207.0236}{\texttt{arXiv:1207.0236}}.

\bibitem{ttxsec5}
\hrefCMSnoop {}{M.~Czakon and A.~Mitov, ``{NNLO corrections to top pair
  production at hadron colliders: the quark-gluon reaction}'',} \textit{ JHEP}
  \textbf{ 01} (2013) 080,
  \href{http://dx.doi.org/10.1007/JHEP01(2013)080}{\doi{10.1007/JHEP01(2013)080}},
\href{http://www.arXiv.org/abs/1210.6832}{\texttt{arXiv:1210.6832}}.

\bibitem{mitov}
\hrefCMSnoop {}{{M. Czakon, P. Fiedler and A. Mitov}, ``{Total Top-Quark
  Pair-Production Cross Section at Hadron Colliders Through
  O($\alpha_S^4$)}'',} \textit{ Phys. Rev. Lett.} \textbf{ 110} (2013) 252004,
  \href{http://dx.doi.org/10.1103/PhysRevLett.110.252004}{\doi{10.1103/PhysRevLett.110.252004}},
\href{http://www.arXiv.org/abs/1303.6254}{\texttt{arXiv:1303.6254}}.

\bibitem{pdf4lhcInterim}
\hrefCMSnoop {}{M.~Botje {et~al.}, ``{The PDF4LHC Working Group Interim
  Recommendations}'',} (2011).
\href{http://www.arXiv.org/abs/1101.0538}{\texttt{arXiv:1101.0538}}.

\bibitem{pdf4lhcReport}
\hrefCMSnoop {}{S.~Alekhin {et~al.}, ``{The PDF4LHC Working Group Interim
  Report}'',} (2011).
\href{http://www.arXiv.org/abs/1101.0536}{\texttt{arXiv:1101.0536}}.

\bibitem{Martin:2009iq}
\hrefCMSnoop {}{A.~D. Martin, W.~J. Stirling, R.~S. Thorne, and G.~Watt,
  ``{Parton distributions for the LHC}'',} \textit{ Eur. Phys. J. C} \textbf{
  63} (2009) 189,
  \href{http://dx.doi.org/10.1140/epjc/s10052-009-1072-5}{\doi{10.1140/epjc/s10052-009-1072-5}},
\href{http://www.arXiv.org/abs/0901.0002}{\texttt{arXiv:0901.0002}}.

\bibitem{mstw08}
\hrefCMSnoop {}{A.~D. Martin, W.~J. Stirling, R.~S. Thorne, and G.~Watt,
  ``{Uncertainties on $\alpha_s$ in global PDF analyses and implications for
  predicted hadronic cross sections}'',} \textit{ Eur. Phys. J. C} \textbf{ 64}
  (2009) 653,
  \href{http://dx.doi.org/10.1140/epjc/s10052-009-1164-2}{\doi{10.1140/epjc/s10052-009-1164-2}},
\href{http://www.arXiv.org/abs/0905.3531}{\texttt{arXiv:0905.3531}}.

\bibitem{Lai:2010vv}
H.-L. Lai\hrefCMSnoop {}{ {et~al.}, ``{New parton distributions for collider
  physics}'',} \textit{ Phys. Rev. D} \textbf{ 82} (2010) 074024,
  \href{http://dx.doi.org/10.1103/PhysRevD.82.074024}{\doi{10.1103/PhysRevD.82.074024}},
\href{http://www.arXiv.org/abs/1007.2241}{\texttt{arXiv:1007.2241}}.

\bibitem{pdfsets}
J.~Gao\hrefCMSnoop {}{ {et~al.}, ``{CT10 next-to-next-to-leading order global
  analysis of QCD}'',} \textit{ Phys. Rev. D} \textbf{ 89} (2014) 033009,
  \href{http://dx.doi.org/10.1103/PhysRevD.89.033009}{\doi{10.1103/PhysRevD.89.033009}},
\href{http://www.arXiv.org/abs/1302.6246}{\texttt{arXiv:1302.6246}}.

\bibitem{Ball:2012cx}
\hrefCMSnoop {}{{NNPDF} Collaboration, ``{Parton distributions with LHC
  data}'',} \textit{ Nucl. Phys. B} \textbf{ 867} (2013) 244,
  \href{http://dx.doi.org/10.1016/j.nuclphysb.2012.10.003}{\doi{10.1016/j.nuclphysb.2012.10.003}},
\href{http://www.arXiv.org/abs/1207.1303}{\texttt{arXiv:1207.1303}}.

\bibitem{PFPAS1}
\href {http://cdsweb.cern.ch/record/1194487}{{CMS Collaboration},
  ``Particle-Flow Event Reconstruction in CMS and Performance for Jets, Taus,
  and MET'',} CMS Physics Analysis Summary CMS-PAS-PFT-09-001, 2009.

\bibitem{PFPAS2}
\href {http://cdsweb.cern.ch/record/1247373}{{CMS Collaboration},
  ``Commissioning of the Particle-flow Event Reconstruction with the first LHC
  collisions recorded in the CMS detector'',} CMS Physics Analysis Summary
  CMS-PAS-PFT-10-001, 2010.

\bibitem{emid}
\hrefCMSnoop {}{{CMS Collaboration}, ``{Performance of electron reconstruction
  and selection with the CMS detector in proton-proton collisions at
  $\sqrt{s}=8\TeV$}'',} \textit{ JINST} \textbf{ 10} (2015) P06005,
  \href{http://dx.doi.org/10.1088/1748-0221/10/06/P06005}{\doi{10.1088/1748-0221/10/06/P06005}},
\href{http://www.arXiv.org/abs/1502.02701}{\texttt{arXiv:1502.02701}}.

\bibitem{muid}
\hrefCMSnoop {}{{CMS Collaboration}, ``{The performance of the CMS muon
  detector in proton-proton collisions at $\sqrt{s} = 7\TeV$ at the LHC}'',}
  \textit{ JINST} \textbf{ 8} (2013) P11002,
  \href{http://dx.doi.org/10.1088/1748-0221/8/11/P11002}{\doi{10.1088/1748-0221/8/11/P11002}},
\href{http://www.arXiv.org/abs/1306.6905}{\texttt{arXiv:1306.6905}}.

\bibitem{inclusWZ3pb}
\hrefCMSnoop {}{{CMS Collaboration}, ``{Measurements of inclusive W and Z cross
  sections in pp collisions at $\sqrt{s} = 7\TeV$}'',} \textit{ JHEP} \textbf{
  01} (2011) 080,
  \href{http://dx.doi.org/10.1007/JHEP01(2011)080}{\doi{10.1007/JHEP01(2011)080}},
  \href{http://www.arXiv.org/abs/1012.2466}{\texttt{arXiv:1012.2466}}.

\bibitem{antikt}
\hrefCMSnoop {}{M.~Cacciari, G.~P. Salam, and G.~Soyez, ``{The anti-$k_{\rm t}$
  jet clustering algorithm}'',} \textit{ JHEP} \textbf{ 04} (2008) 063,
  \href{http://dx.doi.org/10.1088/1126-6708/2008/04/063}{\doi{10.1088/1126-6708/2008/04/063}},
\href{http://www.arXiv.org/abs/0802.1189}{\texttt{arXiv:0802.1189}}.

\bibitem{JESPUB}
\hrefCMSnoop {}{{CMS Collaboration}, ``Determination of jet energy calibration
  and transverse momentum resolution in {CMS}'',} \textit{ JINST} \textbf{ 6}
  (2011) P11002,
  \href{http://dx.doi.org/10.1088/1748-0221/6/11/P11002}{\doi{10.1088/1748-0221/6/11/P11002}},
  \href{http://www.arXiv.org/abs/1107.4277}{\texttt{arXiv:1107.4277}}.

\bibitem{CMStt12}
\hrefCMSnoop {}{{CMS Collaboration}, ``{Measurement of the $\rm t\bar{t}$
  production cross section and the top quark mass in the dilepton channel in
  $\rm pp$ collisions at $\sqrt{s}= 7\TeV$}'',} \textit{ JHEP} \textbf{ 07}
  (2011) 049,
  \href{http://dx.doi.org/10.1007/JHEP07(2011)049}{\doi{10.1007/JHEP07(2011)049}},
\href{http://www.arXiv.org/abs/1105.5661}{\texttt{arXiv:1105.5661}}.

\bibitem{CMStt13}
\hrefCMSnoop {}{{CMS Collaboration}, ``{First measurement of the cross section
  for top-quark pair production in proton-proton collisions at $\sqrt{s}=
  7\TeV$}'',} \textit{ Phys. Lett. B} \textbf{ 695} (2011) 424,
  \href{http://dx.doi.org/10.1016/j.physletb.2010.11.058}{\doi{10.1016/j.physletb.2010.11.058}},
\href{http://www.arXiv.org/abs/1010.5994}{\texttt{arXiv:1010.5994}}.

\bibitem{ttspin_parke}
\hrefCMSnoop {}{G.~Mahlon and S.~J. Parke, ``{Spin correlation effects in top
  quark pair production at the LHC}'',} \textit{ Phys. Rev. D} \textbf{ 81}
  (2010) 074024,
  \href{http://dx.doi.org/10.1103/PhysRevD.81.074024}{\doi{10.1103/PhysRevD.81.074024}},
\href{http://www.arXiv.org/abs/1001.3422}{\texttt{arXiv:1001.3422}}.

\bibitem{ttspin_bernreuther}
\hrefCMSnoop {}{W.~Bernreuther and Z.-G. Si, ``{Top quark spin correlations and
  polarization at the LHC: standard model predictions and effects of anomalous
  top chromo moments}'',} \textit{ Phys. Lett. B} \textbf{ 725} (2013) 115,
  \href{http://dx.doi.org/10.1016/j.physletb.2013.06.051}{\doi{10.1016/j.physletb.2013.06.051}},
  \href{http://www.arXiv.org/abs/1305.2066}{\texttt{arXiv:1305.2066}}.
[Erratum: \DOI{10.1016/j.physletb.2015.03.035}].

\bibitem{ATLASnp3}
\hrefCMSnoop {}{{ATLAS Collaboration}, ``{Measurement of Spin Correlation in
  Top-Antitop Quark Events and Search for Top Squark Pair Production in pp
  Collisions at $\sqrt{s}= 8\TeV$ Using the ATLAS Detector}'',} \textit{ Phys.
  Rev. Lett.} \textbf{ 114} (2015) 142001,
  \href{http://dx.doi.org/10.1103/PhysRevLett.114.142001}{\doi{10.1103/PhysRevLett.114.142001}},
\href{http://www.arXiv.org/abs/1412.4742}{\texttt{arXiv:1412.4742}}.

\bibitem{ATLASspin7}
\hrefCMSnoop {}{{ATLAS Collaboration}, ``{Measurements of spin correlation in
  top-antitop quark events from proton-proton collisions at $\sqrt{s}= 7\TeV$
  using the ATLAS detector}'',} \textit{ Phys. Rev. D} \textbf{ 90} (2014)
  112016,
  \href{http://dx.doi.org/10.1103/PhysRevD.90.112016}{\doi{10.1103/PhysRevD.90.112016}},
\href{http://www.arXiv.org/abs/1407.4314}{\texttt{arXiv:1407.4314}}.

\bibitem{ATLASspinobs}
\hrefCMSnoop {}{{ATLAS Collaboration}, ``{Observation of spin correlation in
  $\rm t \bar{t}$ events from pp collisions at $\sqrt{s} = 7\TeV$ using the
  ATLAS detector}'',} \textit{ Phys. Rev. Lett.} \textbf{ 108} (2012) 212001,
  \href{http://dx.doi.org/10.1103/PhysRevLett.108.212001}{\doi{10.1103/PhysRevLett.108.212001}},
\href{http://www.arXiv.org/abs/1203.4081}{\texttt{arXiv:1203.4081}}.

\bibitem{CMSspin7}
\hrefCMSnoop {}{{CMS Collaboration}, ``{Measurements of $\rm t\bar{t}$ spin
  correlations and top-quark polarization using dilepton final states in $\rm
  pp$ collisions at $\sqrt{s}= 7\TeV$}'',} \textit{ Phys. Rev. Lett.} \textbf{
  112} (2014) 182001,
  \href{http://dx.doi.org/10.1103/PhysRevLett.112.182001}{\doi{10.1103/PhysRevLett.112.182001}},
\href{http://www.arXiv.org/abs/1311.3924}{\texttt{arXiv:1311.3924}}.

\bibitem{lumiPAS13}
\href {http://cdsweb.cern.ch/record/1643269}{{CMS Collaboration}, ``{CMS}
  Luminosity Based on Pixel Cluster Counting - Summer 2013 Update'',} CMS
  Physics Analysis Summary CMS-PAS-LUM-13-001, 2013.

\bibitem{CMSWWZZPublication8}
\hrefCMSnoop {}{{CMS Collaboration}, ``{Measurement of the W$^+$W$^-$ and ZZ
  production cross sections in pp collisions at $\sqrt{s}= 8\TeV$}'',} \textit{
  Phys. Lett. B} \textbf{ 721} (2013) 190,
  \href{http://dx.doi.org/10.1016/j.physletb.2013.03.027}{\doi{10.1016/j.physletb.2013.03.027}},
  \href{http://www.arXiv.org/abs/1301.4698}{\texttt{arXiv:1301.4698}}.

\bibitem{CMSWWPublication7}
\hrefCMSnoop {}{{CMS Collaboration}, ``{Measurement of the W$^+$W$^-$ cross
  section in pp collisions at $\sqrt{s}= 7\TeV$ and limits on anomalous
  WW$\gamma$ and WWZ couplings}'',} \textit{ Eur. Phys. J. C} \textbf{ 73}
  (2013) 2610,
  \href{http://dx.doi.org/10.1140/epjc/s10052-013-2610-8}{\doi{10.1140/epjc/s10052-013-2610-8}},
  \href{http://www.arXiv.org/abs/1306.1126}{\texttt{arXiv:1306.1126}}.

\bibitem{CMSWWHiggsPublication7}
\hrefCMSnoop {}{{CMS Collaboration}, ``{Measurement of W$^+$W$^-$ production
  and search for the Higgs boson in pp collisions at $\sqrt{s} = 7 \TeV$}'',}
  \textit{ Phys. Lett. B} \textbf{ 699} (2011) 25,
  \href{http://dx.doi.org/10.1016/j.physletb.2011.03.056}{\doi{10.1016/j.physletb.2011.03.056}},
  \href{http://www.arXiv.org/abs/1102.5429}{\texttt{arXiv:1102.5429}}.

\bibitem{CMSWWZZPublication7}
\hrefCMSnoop {}{{CMS Collaboration}, ``{Measurement of the sum of WW and WZ
  production with W+dijet events in pp collisions at $\sqrt{s}= 7\TeV$}'',}
  \textit{ Eur. Phys. J. C} \textbf{ 73} (2013) 2283,
  \href{http://dx.doi.org/10.1140/epjc/s10052-013-2283-3}{\doi{10.1140/epjc/s10052-013-2283-3}},
  \href{http://www.arXiv.org/abs/1210.7544}{\texttt{arXiv:1210.7544}}.

\bibitem{CMSZZPublication7}
\hrefCMSnoop {}{{CMS Collaboration}, ``{Measurement of the ZZ production cross
  section and search for anomalous couplings in $2\ell$ $2\ell'$ final states
  in pp collisions at $\sqrt{s}= 7\TeV$}'',} \textit{ JHEP} \textbf{ 01} (2013)
  063,
  \href{http://dx.doi.org/10.1007/JHEP01(2013)063}{\doi{10.1007/JHEP01(2013)063}},
  \href{http://www.arXiv.org/abs/1211.4890}{\texttt{arXiv:1211.4890}}.

\bibitem{CMStopPublicationstop4}
\hrefCMSnoop {}{{CMS Collaboration}, ``{Measurement of the single-top-quark
  $t$-channel cross section in pp collisions at $\sqrt{s}= 7\TeV$}'',} \textit{
  JHEP} \textbf{ 12} (2012) 035,
  \href{http://dx.doi.org/10.1007/JHEP12(2012)035}{\doi{10.1007/JHEP12(2012)035}},
  \href{http://www.arXiv.org/abs/1209.4533}{\texttt{arXiv:1209.4533}}.

\bibitem{ATLASWWPublication}
\hrefCMSnoop {}{{ATLAS Collaboration}, ``{Measurement of the WW cross section
  in $\sqrt{s}= 7\TeV$ pp collisions with the ATLAS detector and limits on
  anomalous gauge couplings}'',} \textit{ Phys. Lett. B} \textbf{ 712} (2012)
  289,
  \href{http://dx.doi.org/10.1016/j.physletb.2012.05.003}{\doi{10.1016/j.physletb.2012.05.003}},
  \href{http://www.arXiv.org/abs/1203.6232}{\texttt{arXiv:1203.6232}}.

\bibitem{ATLASWZPublication}
\hrefCMSnoop {}{{ATLAS Collaboration}, ``{Measurement of the W$^{\pm}$Z
  production cross section and limits on anomalous triple gauge couplings in
  proton-proton collisions at $\sqrt{s}= 7\TeV$ with the ATLAS detector}'',}
  \textit{ Phys. Lett. B} \textbf{ 709} (2012) 341,
  \href{http://dx.doi.org/10.1016/j.physletb.2012.02.053}{\doi{10.1016/j.physletb.2012.02.053}},
  \href{http://www.arXiv.org/abs/1111.5570}{\texttt{arXiv:1111.5570}}.

\bibitem{ATLASZZPublication}
\hrefCMSnoop {}{{ATLAS Collaboration}, ``{Measurement of the $\rm ZZ$
  Production Cross Section and Limits on Anomalous Neutral Triple Gauge
  Couplings in Proton-Proton Collisions at $\sqrt{s}= 7\TeV$ with the ATLAS
  Detector}'',} \textit{ Phys. Rev. Lett.} \textbf{ 108} (2012) 041804,
  \href{http://dx.doi.org/10.1103/PhysRevLett.108.041804}{\doi{10.1103/PhysRevLett.108.041804}},
  \href{http://www.arXiv.org/abs/1110.5016}{\texttt{arXiv:1110.5016}}.

\bibitem{BTV-11-004-pub}
\hrefCMSnoop {}{{CMS Collaboration}, ``Identification of b-quark jets with the
  {CMS} experiment'',} \textit{ JINST} \textbf{ 8} (2013) P04013,
  \href{http://dx.doi.org/10.1088/1748-0221/8/04/P04013}{\doi{10.1088/1748-0221/8/04/P04013}},
  \href{http://www.arXiv.org/abs/1211.4462}{\texttt{arXiv:1211.4462}}.

\bibitem{nuisance}
\hrefCMSnoop {}{D.~A.~S. Fraser, N.~Reid, and A.~C.~M. Wong, ``{Inference for
  bounded parameters}'',} \textit{ Phys. Rev. D} \textbf{ 69} (2004) 033002,
\href{http://dx.doi.org/10.1103/PhysRevD.69.033002}{\doi{10.1103/PhysRevD.69.033002}}.

\bibitem{tevatron_combi}
\hrefCMSnoop {}{{CDF and D0 Collaborations}, ``{Combination of measurements of
  the top-quark pair production cross section from the Tevatron Collider}'',}
  \textit{ Phys. Rev. D} \textbf{ 89} (2014) 072001,
  \href{http://dx.doi.org/10.1103/PhysRevD.89.072001}{\doi{10.1103/PhysRevD.89.072001}},
\href{http://www.arXiv.org/abs/1309.7570}{\texttt{arXiv:1309.7570}}.

\end{thebibliography}\endgroup
\numberwithin{figure}{section}
\ifthenelse{\boolean{cms@external}}{}{
\clearpage
\appendix
\section{\texorpdfstring{\ttbar production cross section in $\Pp\Pap$ and $\Pp\Pp$ collisions as a function of $\sqrt{s}$}{ttbar production cross section in proton-antiproton and proton-proton collisions as a function of sqrt(s)}\label{app:supp_material}}
\begin{figure*}[htbp]
\centering
\includegraphics[width=0.98\textwidth]{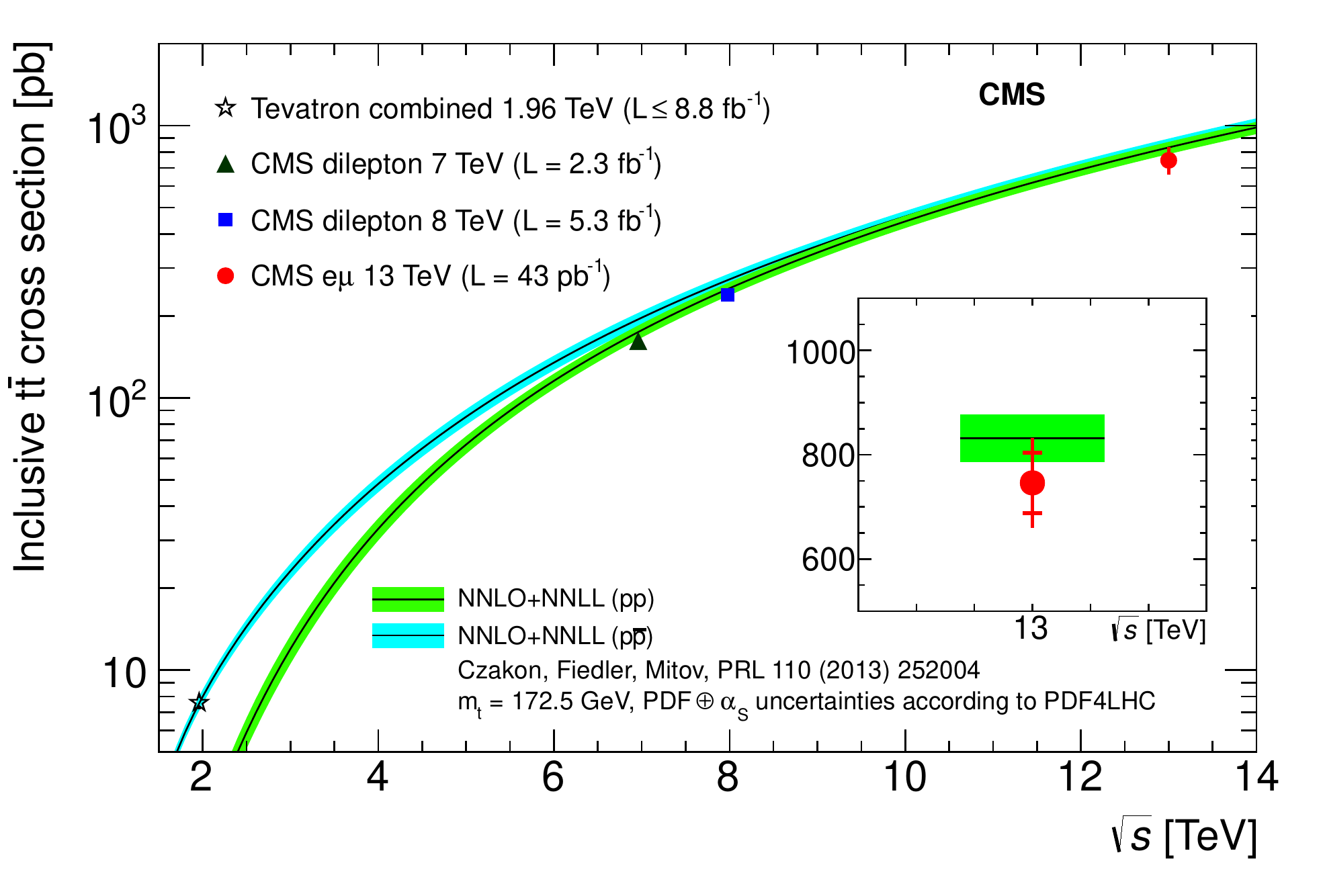}
\caption{The \ttbar production cross section in $\Pp\Pap$ and $\Pp\Pp$ collisions as a function of $\sqrt{s}$. The Tevatron
combination is given at $\sqrt{s}=1.96\TeV$~\cite{tevatron_combi}, as
are the CMS results at 7 and 8\TeV in the dilepton
channels~\cite{CMStt3,CMStt8}. The CMS result at 13\TeV is also
shown in the figure where the inner error bar corresponds to the
statistical uncertainty and the outer one to the total
uncertainty. The measurements are compared to NNLO+NNLL theoretical
predictions~\cite{mitov}.}
\label{fig:sqrt}
\end{figure*}

}

\cleardoublepage \section{The CMS Collaboration \label{app:collab}}\begin{sloppypar}\hyphenpenalty=5000\widowpenalty=500\clubpenalty=5000\textbf{Yerevan Physics Institute,  Yerevan,  Armenia}\\*[0pt]
V.~Khachatryan, A.M.~Sirunyan, A.~Tumasyan
\vskip\cmsinstskip
\textbf{Institut f\"{u}r Hochenergiephysik der OeAW,  Wien,  Austria}\\*[0pt]
W.~Adam, E.~Asilar, T.~Bergauer, J.~Brandstetter, E.~Brondolin, M.~Dragicevic, J.~Er\"{o}, M.~Flechl, M.~Friedl, R.~Fr\"{u}hwirth\cmsAuthorMark{1}, V.M.~Ghete, C.~Hartl, N.~H\"{o}rmann, J.~Hrubec, M.~Jeitler\cmsAuthorMark{1}, V.~Kn\"{u}nz, A.~K\"{o}nig, M.~Krammer\cmsAuthorMark{1}, I.~Kr\"{a}tschmer, D.~Liko, T.~Matsushita, I.~Mikulec, D.~Rabady\cmsAuthorMark{2}, B.~Rahbaran, H.~Rohringer, J.~Schieck\cmsAuthorMark{1}, R.~Sch\"{o}fbeck, J.~Strauss, W.~Treberer-Treberspurg, W.~Waltenberger, C.-E.~Wulz\cmsAuthorMark{1}
% ...
\vskip\cmsinstskip
% ...
1:~~Also at Vienna University of Technology, Vienna, Austria\\
2:~~Also at CERN, European Organization for Nuclear Research, Geneva, Switzerland\\
% ...

\end{sloppypar}
\end{document}